\documentclass[12pt]{iopart}

\usepackage{bm}
\usepackage{graphicx}
\usepackage{subfigure}
\begin{document}

\title{Coupling-Matrix Approach to the Chern Number Calculation in Disordered Systems}
\author{Y. F. Zhang$^1$, Y. Y. Yang$^1$, Yan Ju$^1$, L. Sheng$^{*1}$, R. Shen$^1$, D. N. Sheng$^2$, D. Y. Xing$^{\dag 1}$}
\address{$^1$National Laboratory of Solid State Microstructures and
Department of Physics, Nanjing University, Nanjing 210093, China\\
$^2$Department of Physics and Astronomy, California State
University, Northridge, California 91330, USA}
\ead{\mailto{$*$ shengli@nju.edu.cn}, \mailto{$\dag$ dyxing@nju.edu.cn}}

\begin{abstract}
The Chern number is often used to distinguish between different
topological phases of matter in two-dimensional electron systems.
A fast and efficient coupling-matrix
method is designed to calculate the Chern number
in finite crystalline and disordered systems. To show
its effectiveness, we apply the approach to the Haldane model
and the lattice Hofstadter model, the
quantized Chern numbers being correctly obtained.
The disorder-induced topological phase transition is well
reproduced, when the disorder strength is increased beyond
the critical value. We expect the method to be widely applicable
to the study of topological quantum numbers.

% While the traditional ways
% to calculate Chern number is time-consuming, especially for disorder system.
% In this paper, we addressed a new Chern number formula for disordered
% topological system. The new way has the advantage of being easier
%  and time saving to calculate. We apply this formula to Chern insulaltor and lattice
%  Hofstadter model, and show that a disorder induced topological phase
%  transition occurs when the disorder strength is increased beyond a critical value.
\end{abstract}

\pacs{73.43.Nq, 71.23.An,  72.80.Vp}
\maketitle

\section{Introduction}
%\text{\emph{Introduction}.-}
In the past thirty years, the condensed matter physics community has
been fascinated by topological phases of matter, for instance, the
integer quantum Hall effect,~\cite{iqhe} the fractional quantum Hall
effect,~\cite{fqhe} the quantum anomalous Hall effect,~\cite{ci}  the
quantum spin Hall effect,~\cite{qshe1,qshe2} and the
three-dimensional topological insulators.~\cite{3d1,3d2} These
topological states of quantum matter are usually distinguished by
use of some global topological quantum numbers~\cite{toponum} rather
than certain local order parameters. The topological aspect of the
integer quantum Hall effect with periodic potentials was first
discussed by Thouless, Kohmoto, Nightingale, and Nijs
(TKNN).~\cite{tknn} In their famous work, a topological expression
for the Hall conductivity was given by the Chern number over the
magnetic Brillouin zone. Their result was then generalized to the
fractional quantum Hall effect.~\cite{fchern} For the
quantum spin Hall systems, with the extension of the idea, the
well-defined spin Chern number can be used to characterize  trivial
and non-trivial bulk band topology.~\cite{spinch1,spinch2}

While simplification exists for pure systems~\cite{Hatsugai},
calculation of the Chern number in the presence of disorder is
usually based upon the integral of partial derivatives of electron
wave functions over the boundary phases.~\cite{fchern,Bhatt,disorder1}
Numerical implementation involves hundreds of times of exact
diagonalization for a given disorder
configuration~\cite{Bhatt,disorder1,dnsheng,Bhatt1}, which is very time-consuming
even for noninteracting electron systems. Recently, several
different approaches for the Chern number computation in real space
have been suggested in the literature. Kitaev~\cite{trace} proposed
a real space Chern number formula for a lattice model in terms of
traces of the coordinate operator and projection operator. A recent
work similar to Kitaev's one was numerically realized by Bianco and
Resta~\cite{Resta}, and they generalized the result to the systems
with open boundary conditions. In the presence of disorder, however,
the idea of supercells,~\cite{supercell} namely, a periodic
duplication of the actual system is needed, which greatly increases
the computation time. Based upon the noncommutative Chern number
theory, Prodan $et$ $al.$~\cite{prodanPRL} proposed an efficient
method to calculate the Chern number of disordered systems, in which
the procedure of the exact diagonalization is greatly simplified.
However, it involves somewhat complicated multiple commutators
between the coordinate operators and projection operators.
In some cases, the Chern number can be extracted indirectly from
the transport coefficients, which can be relatively easily calculated from
the Kubo formula~\cite{dnsheng}.
In some other cases, direct calculation of the topological
number of bulk wavefunctions is often needed and sometimes
irreplaceable. Therefore, development of efficient
numerical approaches to direct calculation of
the Chern number is highly desirable.

In this work, we provide an alternative way to calculate the Chern
number, in which only one time exact diagonalization for the actual
system is needed without loss of accuracy. A transparent coupling-matrix
formulation will be given, from which the Chern number can be very
efficiently computed,  compared with the existing approaches. To show
its effectiveness,  this approach is applied to both the Haldane
model and the Hofstadter model. The calculated Chern number is found
well quantized provided the Fermi level lies within the energy gap,
even when the sample size is not very large. The topological
phase transition from the quantum Hall insulator to an ordinary insulator
can be determined based upon the calculated Chern number, and the
obtained critical disorder strength is in good agreement with the result
previously obtained from the Hall conductivity calculation.

In the next section, we present the new approach of calculating the
Chern number. In Sec.\ III, for the Haldane model, it is
shown that the present approach works well for both crystalline and
disordered systems. In Sec.\ IV, we apply the approach to the
lattice Hofstadter model and the calculated results are in
accordance with already existing results. The final section is a
summary.

\section{Method}

We now consider a two dimensional (2D) lattice with
$N=L_{x}\times L_{y}$ unit cells. We use ${\bf r}=(x,y)$
with $x$ and $y$ as integers to index the position of a unit cell.
We can define twisted boundary conditions for a single-particle
wavefunction of the system
$\varphi_{\theta}(x+L_{x},y)=e^{i\theta_{x}}\varphi_{\theta}(x,y)$ and
$\varphi_{\theta}(x,y+L_{y})=e^{i\theta_{y}}\varphi_{\theta}(x,y)$ with $\theta=(\theta_x,\theta_y)$
and $0\leq\theta_{x},\theta_{y}\leq2\pi$. The wavefunction is in general
a row vector in the space of inner degrees of freedom, such as
spin and sublattice. The system is assumed to have $M$ electrons, and in the ground state
the wavefunctions of the $M$ occupied single-particle electron states
are denoted by $\varphi_{\theta}^{m}(\mathbf{r})$ with
$m=0,1,\cdots M-1$. The many-body
wavefunction of the ground state $\Psi_{\theta}(\{{\bf r}_i\})$
is the Slater determinant of the single-particle
wavefunctions $\varphi_{\theta}^{m}(\mathbf{r}_{i})$,
where ${\bf r}_{i}$ with $i=0,1,\cdots M-1$ is the coordinate
of the $i$-th electron.
The Chern number of the gound state is given
by~\cite{fchern,disorder1,Bhatt}
\begin{equation}\label{old}
C=\frac{1}{2\pi i} \int_{T_{\theta}} d{\theta}
\langle\nabla_{\theta}\Psi_{\theta}|\times
|\nabla_{\theta}\Psi_{\theta}\rangle\ ,
\end{equation}
with $\mathbf{\theta}=(\theta_{x},\theta_{y})$ and
$T_{{\mathbf{\theta}}}$ denoting the
$(\theta_{x},\theta_{y})$ space, which is essentially a torus.

We transform the single-particle eigenstate from real space
to momentum space through a Fourier transform (FT)
\begin{equation}\label{ft}
\varphi_{\theta}^{m}(\mathbf{r}_{i})= \frac{1}{\sqrt{N}}\sum_{\mathbf{k}_{i}}
F^m(\mathbf{k}_{i})e^{i{\mathbf{k}_{i}}\cdot{\bf r}_{i}}\ .
\end{equation}
The twisted boundary conditions require that the
momenta take only the discrete values
$\mathbf{k}_{i}={\bf k}^{(0)}_{i}+{\bf q}$, where
$\mathbf{k}^{(0)}_{i}=(\frac{2n\pi}{L_{x}},\frac{2l\pi}{L_{y}})$ with $0\leq
n < L_{x}$ and $0\leq l < L_{y}$, and
$\mathbf{q}=(\frac{\theta_{x}}{L_{x}},\frac{\theta_{y}}{L_{y}})$. We
note that the set of $\{\mathbf{k}^{(0)}_{i}\}$ are actually the discrete momenta for
periodic boundary conditions. We will denote $F^{m}({\bf k}_{i})\equiv F^{m}({\bf k}_{i}^{(0)}
+{\bf q})$ as
$F_{\bf q}^{m}({\bf k}^{(0)}_{i})$. It is easy to find
that the many-body wavefunction of the ground state in the momentum space
$\Phi_{\bf q}(\{\mathbf{k}^{(0)}_{i}\})$ is the Slater determinant
of $F_{\bf q}^{m}(\mathbf{k}^{(0)}_{i})$.
By means of the substitution
$\frac{\partial}{\partial \theta_{x}}\rightarrow
\frac{1}{L_{x}}\frac{\partial}{\partial q_{x}}$ and $\frac{\partial}{\partial \theta_{y}}\rightarrow
\frac{1}{L_{y}}\frac{\partial}{\partial q_{y}}$, the Chern number
is derived to be
\begin{eqnarray}\label{new}
C=\frac{1}{2\pi i}\int_{R_{\mathbf{q}}}d{\bf q}\langle
\nabla_{\mathbf{q}} \Phi_{\bf q}\vert
\times\vert
\nabla_{\mathbf{q}} \Phi_{\mathbf{q}}\rangle
\end{eqnarray}
where the inner product includes a summation over $\{{\bf k}^{(0)}\}$,
and $R_{\mathbf{q}}$ denotes the rectangle of $[0,
\frac{2\pi}{L_{x}}]\times[0,\frac{2\pi}{L_{y}}]$.

%Substituting
%$\mathbf{k=k^0+q}$ and $\nabla_{\mathbf{q}}\rightarrow
%\nabla_{\mathbf{k}}$ into Eq.\ (\ref{new}), we obtain
%\begin{equation}\label{newk}
%  C_{m}=\frac{i}{2\pi}\int_{BZ}
%d\mathbf{k} \nabla_{\mathbf{k}} F^{*}_{m}(\mathbf{k}) \times
%\nabla_{\mathbf{k}} F_m(\mathbf{k})\ .
%\end{equation}
%The Chern number Eq.\ (\ref{newk}) becomes an integral over the Brillouin
%zone $[0, 2\pi]\times[0, 2\pi]$ shown in Fig.\ 1(a). This
%formula has prevously been obtained by Fukui, Hatsugai and Suzuki~\cite{jpsj}
%for clean systems.
%These authors also showed that it can precisely produce
%the correct quantized Chern number when the size of the system
%$N>N^{c}$ with  $N^{c}\approx \mathcal{O}(|C_{m}|)$ as a critical
%size.

\begin{figure}
 \centering

    \includegraphics[width=4in]{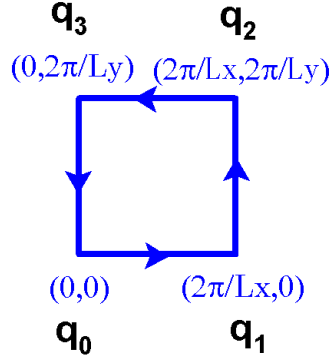}

\caption{The rectangle $R_{\mathbf{q}}$ has a size
$4\pi^{2}/L_{x}L_{y}$, whose four vertices are denoted by ${\bf
q}_\alpha (\alpha=0,1,2,3$ ).}
\label{geo} % label for entire figure
\end{figure}
Using the Stokes theorem, we rewrite Eq.\ (\ref{new}) as a line
integral
\begin{equation}\label{new2}
C=\frac{1}{2\pi i}\oint_{\partial R_{\mathbf{q}}}d{\bf l}_q\cdot
\langle
\Phi_{\bf q}|\nabla_{{\bf q}}
   \Phi_{\bf q}\rangle\ ,
\end{equation}
around the boundary of $R_{\mathbf{q}}$, denoted by $\partial
R_{\mathbf{q}}$. The Chern number given in Eq.\ (\ref{new2}) is expressed
as a winding number along closed path $\partial R_{\mathbf{q}}$.
We can divide $\partial R_{\mathbf{q}}$ into $N$ small line segments
with ${\bf q}_{\alpha}$ ($\alpha=0,1\cdots N$) as their endpoints.
In each segment, one can replace the derivatives in Eq.\
(\ref{new2}) by discrete differences and the
integral by a summation, yielding
\begin{equation}\label{newq}
C=\frac{1}{2\pi}\sum_{\alpha=0}^{N-1}\mbox{Arg}[
\mbox{det}(C_{\alpha,\alpha+1})]\ ,
\end{equation}
where $\mbox{Arg}(\cdot)$ stands for the principal argument, and
$C_{\alpha,\alpha+1}$ is a $M\times M$ coupling matrix,~\cite{Baer}
with elements $C^{mn}_{\alpha,\alpha+1}=\langle
F^{m}_{{\bf q}_\alpha}|F^{n}_{{\bf
q}_{\alpha+1}}\rangle$. Here, we have used the relation $\langle\Phi_{{\bf q}_{\alpha}}
\vert\Phi_{{\bf q}_{\alpha+1}}\rangle=det(C_{\alpha,\alpha+1})$.
 Equivalently, one
can first multiply the coupling matrices $\tilde{C}=\prod_{\alpha=0}^{N-1}C_{\alpha,\alpha+1}$,
and then diagonalize $\tilde{C}$. The Chern number is given by the
sum of the phases of the eigenvalues of $\tilde{C}$ divided by $2\pi$.

In practice, it is sufficient to take $N=4$ and ${\bf q}_{\alpha}$
with $\alpha=0,1,2,$ and $3$ (${\bf q}_4={\bf q}_0$) to be the four
vertices of $R_{\mathbf{q}}$ shown in Fig.\ \ref{geo},
when $L_{x}\gg 1$ and $L_{y}\gg 1$. It is interesting to notice that for these ${\bf q}_\alpha$,
${\bf k}^{(0)}+{\bf q}_\alpha$ still belong
to the set of $\{{\bf k}^{(0)}\}$, so all the quantities
that are needed for the calculation of the Chern number
can be obtained in the system with periodic
boundary conditions. Moreover, using the inverse FT, the matrix
elements of the coupling matrix $C_{\alpha,\alpha+1}$
can be expressed in real space as
\begin{equation}\label{coupling}
C^{mn}_{\alpha,\alpha+1}=\langle\varphi_{\theta=0}^m\vert e^{i({\bf q}_{\alpha}
-{\bf q}_{\alpha+1})\cdot{\bf r}}\vert\varphi_{\theta=0}^n\rangle\ .
\end{equation}
Here, $\varphi_{\theta=0}^m({\bf r})$ with $m=0,1,2\cdots(M-1)$
are the single-particle wave functions of the occupied electron
states for periodic boundary conditions.
The product of the coupling matrices
\begin{equation}
\widetilde{C}=C_{0,1}C_{1,2}C_{2,3}C_{3,0}\ ,
\end{equation}
is first diagonalized to find
$M$ eigenvalues denoted as $\lambda_m$. Since $\widetilde{C}$
is not Hamiltian, its eigenvalues $\lambda_m$ are usually complex numbers.
Then the Chern number is given by
\begin{equation}
C=\frac{1}{2\pi}\sum_{m=0}^{M-1}\mbox{Arg}(\lambda_m)\ ,
\end{equation}
where the range of the principal argument
function $\mbox{Arg}(\cdot)$ is taken to be $(-\pi, \pi]$.
We mention that $\widetilde{C}$ tends to be diagonal when
$M$ is sufficiently large~\cite{Baer}, so the sum of the phases of all the
diagonal elements in $\widetilde{C}$ divided by $2\pi$
yields a good approximation to
the Chern number.

Prodan, Hughes, and Bernevig~\cite{prodanPRL} proposed an
efficient numerical approach to
the Chern number, by extending the formula for pure systems straightforwardly
to disordered systems. Their expression is also in real space,
which however involves multiple commutators of the coordinate operators
and projection operator, and is unlikely reducible to a single term.
Our method has the advantage that
the formula contains only a single term
and is more physically transparent. In the next two sections, numerical
calculations will be carried out to show the validity and efficiency
of this formula for calculating the Chern number.

%Here we stress that the special case of $M=1$, our method is equivalent to the
%following Chern number formula ( ):
%\begin{equation}\label{newkk}
%  C_{m}=\frac{i}{2\pi}\int_{BZ}
%d\mathbf{k} \nabla_{\mathbf{k}} F^{*}_{m}(\mathbf{k}) \times \nabla_{\mathbf{k}} F_m(\mathbf{k})
%\end{equation}
%where $F_m(\mathbf{k})$ is the Fourier transformation of single-particle eigenstate ( see also Appendix ).
%Chern number (\ref{newkk}) is a internal on Brillouin zone $[0, 2\pi]\times[0, 2\pi]$, no
%matter whether the system is disordered. It is notable that, for a system without disorder,
%Chern number (\ref{newkk}) is the same as the TKNN Chern number.

\section{The Haldane Model}
Let us consider the Haldane model~\cite{ci} for the quantum anomalous Hall effect
defined on a honeycomb lattice, including a random on-site disorder
potential
\begin{equation}
H=-\sum_{\langle i,j\rangle}\hat{c}_i^\dag\hat{c}_j+it\sum_{\langle\langle
i,j\rangle\rangle}v_{ij}\hat{c}_i^\dag\hat{c}_j+\sum_{i}\omega_{i}\hat{c}_i^\dag\hat{c}_i\
.
\end{equation}
Here, the first term describes the usual nearest-neighbor hopping
with the hopping integral being taken to be unity,
and the second term stands for the next-nearest-neighbor hopping
with a complex hopping integral and $v_{ij}=(\bm{d}_{kj}\times
\bm{d}_{ik})_{z}/|(\bm{d}_{kj}\times \bm{d}_{ik})_{z}|$. $i$ and $j$
are two next nearest neighbor sites and $k$ their common nearest
neighbor. The vector $\bm{d}_{ik}$ points from $k$ to $i$, and
the distance between two nearest neighbor sites is taken to be
unity. The third term represents a random on-site potential, with
$\omega_{i}$ uniformly distributed between [$W/2$, $W/2$]. It is
known~\cite{ci} that in the absence of disorder, the energy spectrum
of the Haldane model has a middle band gap between $-3\sqrt{3}t$ and $3\sqrt{3}t$.
When the Fermi energy lies in the band gap, the system shows a
quantized Hall conductivity ${e^2}/{h}$, without an applied magnetic
field.
\begin{figure}
 \centering
\includegraphics[width=4in]{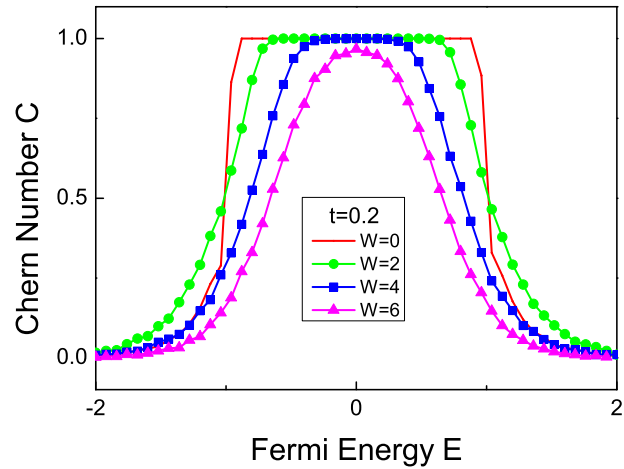}
\caption{ Chern number for the Haldane model as a function of
electron Fermi energy $E$ for $t=0.2$, for several different disorder strengths.
The result is averaged over 400 disorder
configurations for sample size $N=48\times 48$. }
\label{haldane1}
\end{figure}

\begin{figure}[!htb]
 \centering
\includegraphics[width=4in]{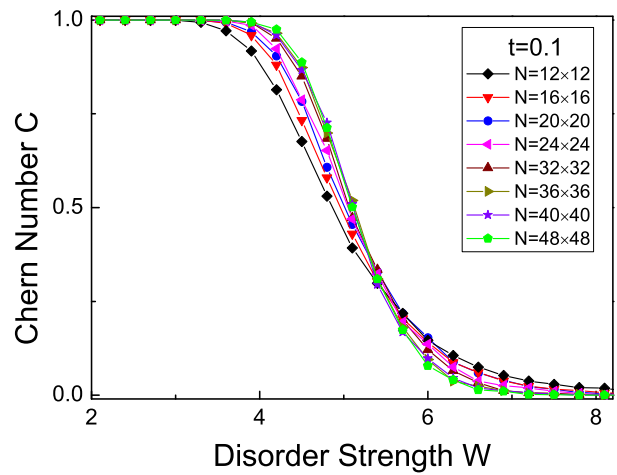}
\caption{ Chern number for the Haldane model
as a function of disorder strength $W$ at $t=0.1$, for several different
sample sizes. The result is averaged over 400 disorder
configurations.}
\label{haldane3}
\end{figure}

Numerical calculation of the Chern number is carried out for a
disordered system with size N=$48\times48$ and $t=0.2$, and the
results are shown in Fig.\ \ref{haldane1}. In the absence of
disorder ($W=0$), the Chern number is well quantized to 1 within the
middle band gap, which is consistent with the known
result~\cite{ci}. With increasing disorder strength, the $C=1$
plateau narrows, manifesting the levitation and pair
annihilation~\cite{nagaosa} of extended states for the valence and
conduction bands, whose Chern numbers have opposite signs. The
calculated Chern number at a fixed Fermi energy ($E=0$) for $t=0.1$
as a function of disorder strength is plotted in Fig.\
\ref{haldane3} for different sample sizes. It is found that the
Chern number is robust against weak disorder $W<4$. With increasing
$W$ from 4 to 6, the Chern number continuously decreases to nearly
zero. With increasing the sample size, the transition process
becomes sharper and sharper, which conforms the expectation that the
transition should become a sudden drop from 1 to 0 in the
thermodynamic limit.

\section{Hofstadter model}

\begin{figure}[!htb]
 \centering
\includegraphics[width=4in]{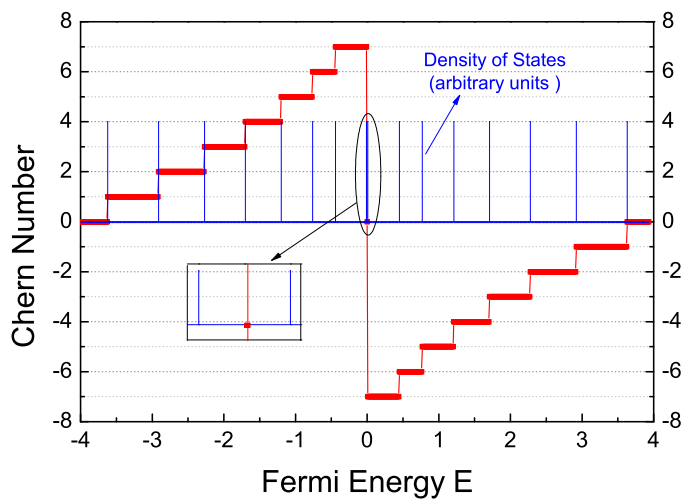}
\caption{ Calculated Chern number and electron
density of states for the lattice Hofstadter model in the full
energy band for magnetic flux $q/p=1/16$. The disorder strength is
set to be $W=0$ and $N=64 \times 64$. The inset is a drawing of partial
enlargement.} \label{flat}
\end{figure}

We next turn to the Hofstadter model~\cite{hofs} on a 2D
square lattice. Under an external uniform magnetic field with the
Landau gauge $\vec{A}=(yB,0,0)$, the model Hamiltonian is given by
\begin{equation}\label{hof}
H_{0}=-\sum\limits_{m,n}(e^{in\phi}c^{\dag}_{m,n}c_{m+1,n}+c^{\dag}_{m,n}c_{m,n+1})+H.c.,
\end{equation}
where integers $m$ and $n$ are the $x$ and $y$ coordinates of a
lattice site, $c_{m,n}$ is the fermion annihilation operator on the
site, and the hopping integral has been set to be the unit of
energy. The magnetic flux quanta per plaquette is $\phi=2\pi q/p$.
In the absence of disorder, the original band splits into $p$ Landau
levels for mutually prime integers $p$ and $q$. The special case of
$q=1$ has been widely studied in previous
works.~\cite{disorder1,dnsheng,Bhatt1,hof1,hof2,hof4} It was found that each Landau
band carries a Chern number $+1$ except the band at the center
($E=0$) which carries a negative Chern number $-(p-1)$ for odd $p$
or $-(p/2-1)$ for even $p$. Figure\ \ref{flat} shows the calculated
Chern number as a function of the Fermi energy at $q/p=1/16$ for a
sample of size N=$64 \times 64$. We see that the Chern number of
each Landau level is $+1$, except the central two Landau levels,
each one having a Chern number $-7$. As a result, the sum of the
Chern numbers of all the Landau levels in the full energy band is
zero.

\begin{figure}
\centering
\includegraphics[width=4in]{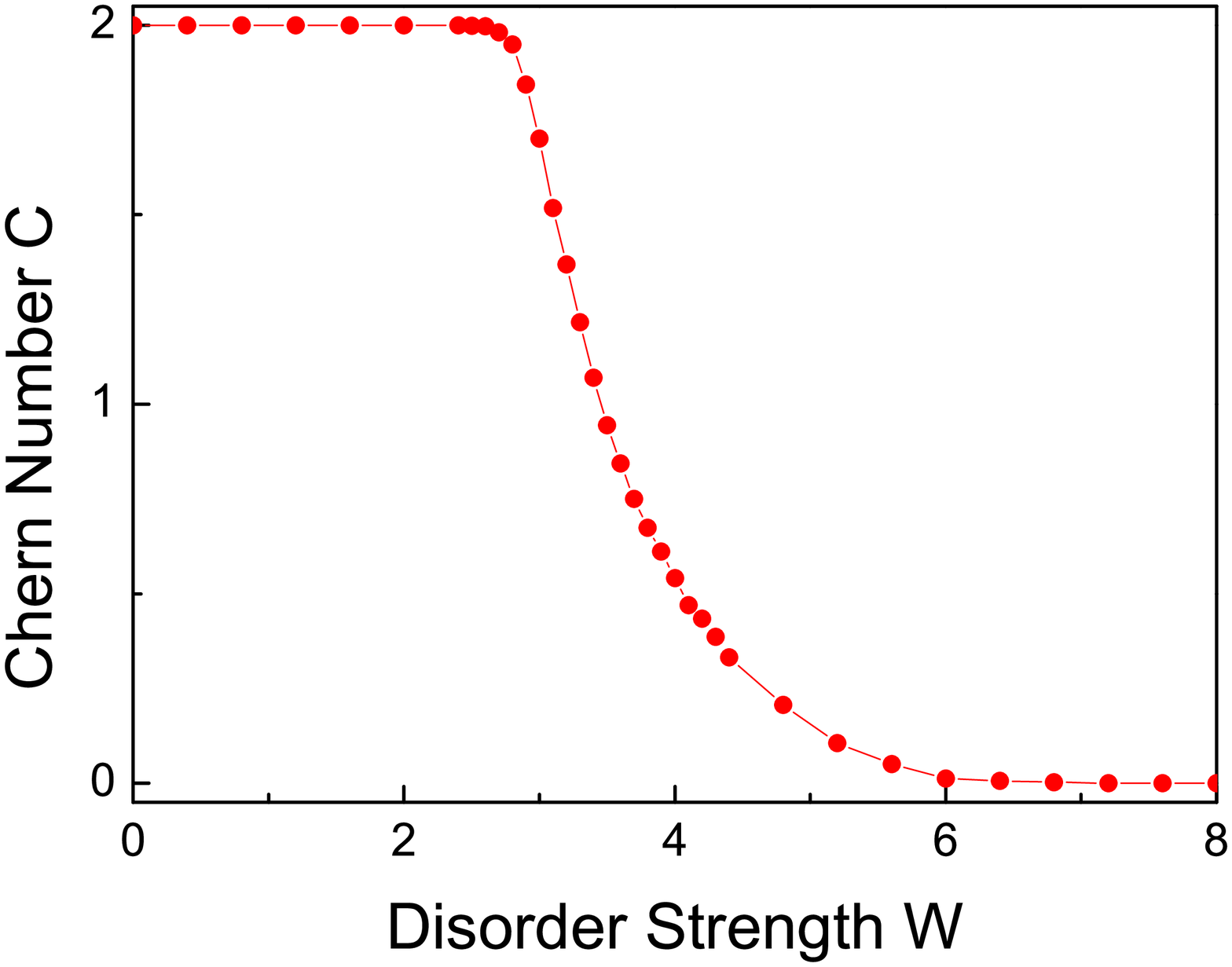}
\caption{ Chern number as a function of
disorder strength $W$ for the
lattice Hofstadter for magnetic flux
$q/p=1/16$ and sample size $N=64 \times
64$. The result is averaged over 400 disorder configurations.} \label{disorder}
\end{figure}

To study the disorder effect, we include into the Hamiltonian a
random on-site potential of the form $\sum_{m,n}\omega_{m,n}
c^{\dag}_{m,n}c_{m,n}$, with $w_{m,n}$ randomly distributed between
$[-W/2,W/2]$. The calculated Chern number at a fixed Fermi energy
$E=-2.75$ is displayed as a function of disorder strength $W$ in
Fig.\ \ref{disorder}. The Chern number is well quantized for weak
disorder. With increasing $W$,  the interesting direct transition~\cite{dnsheng}
from the $C=2$ quantum Hall plateau to a trivial insulator state
with $C=0$ is reproduced. Moreover,
the critical disorder strength is
estimated to be about $W=3.5$, in good agreement with the result obtained from the
Kubo formula calculation~\cite{dnsheng}.

%\begin{figure}
%\includegraphics[width=3.6in]{haldane1.eps}
%\caption{ The disorder averaged Chern number for Chern insulator at $E_{F}$ =0 as a function of the disorder
%strength W for parameter t=0.05, 0.1, 0.2.
%The size is set to $N=48 \times 48$ and disorder is averaged over 400 times.}
%\label{haldane2}
%\end{figure}
\section{ Summary}
To conclude, we have proposed an efficient coupling-matrix method
for calculating the Chern number of disordered systems.  The present
approach is applied to both the Haldane model and lattice Hofstadter
model. The calculated Chern number is found well quantized  even if
the sample size is not very large. The calculated results, in
particular the disorder-induced phase transition, are in good
agreement with the known results. Our approach can be directly applied to
calculate the spin Chern numbers~\cite{spinch1,spinch2} for the quantum spin Hall systems,
from which the Z$_{2}$ index can be extracted.

\section{ACKNOWLEDGMENTS}
This work is supported by the State Key Program for Basic Researches
of China under Grants Nos. 2009CB929504 (LS),
2011CB922103, and 2010CB923400 (DYX), the National Natural Science
Foundation of China under Grant Nos. 11225420, 11074110 (LS), 11074111 (RS),
11174125, 11074109, 91021003 (DYX), Natural Science
Foundation of Jiangsu Province in China under grant No. BK2010364 (YJ),
and a project funded by the PAPD
of Jiangsu Higher Education Institutions. We also thank the US NSF Grants
No. DMR-0906816 and No. DMR-1205734, and
Princeton MRSEC Grant No.DMR-0819860 (DNS).

\end{document}